\documentclass[nohyper,12pt,letterpaper]{JHEP}  
\usepackage{epsfig}

\newfont{\frak}{eufm10 scaled 1200}

\newfont{\Bbb}{msbm10 scaled 1200}     
\newcommand{\mathbb}[1]{\mbox{\Bbb #1}}
\DeclareSymbolFont{AMSa}{U}{msa}{m}{n}
\DeclareSymbolFont{AMSb}{U}{msb}{m}{n}
\let\Box\relax
\DeclareMathSymbol{\Box}{\mathord}{AMSa}{"03}

\def\IR{{\mathbb R}}

\def\IQ{{\mathbb Q}}

\def \eqn#1#2{\begin{equation}#2\label{#1}\end{equation}} 
\def \rut{2/5 transformation}
\def \Rut{{\mathbb G}}             
\def \lpl{L_{planck}}
\def \lst{L_{string}}

\def \Lag{\mathcal L}

\def\hacek{\accent20}                           

\title{Dualities versus Singularities}

\author{Tom Banks\\
  Department of Physics and Astronomy\\
  Rutgers University, Piscataway, NJ 08855-0849\\
E-mail: \email{banks@physics.rutgers.edu}}

\author{Willy Fischler\\
  Theory Group, Department of Physics\\
  University of Texas, Austin, TX 78712\\
E-mail: \email{fischler@physics.utexas.edu}}

\author{Lubo\hacek s Motl\\
  Department of Physics and Astronomy\\
  Rutgers University, Piscataway, NJ 08855-0849\\
E-mail: \email{motl@physics.rutgers.edu}}

\abstract{
We show that a subgroup of the modular group of M-theory compactified on a
ten torus, implies the Lorentzian structure of the moduli space, that is
usually associated with naive discussions of quantum cosmology based on
the low energy Einstein action. This structure implies a natural division
of the asymptotic domains of the moduli space into regions which
can/cannot be mapped to Type II string theory or 11D Supergravity (SUGRA)
with large radii.  We call these the safe and unsafe domains. The safe
domain is the interior of the future light cone in the moduli space while
the unsafe domain contains the spacelike region and the past light cone.
Within the safe domain, apparent cosmological singularities can be
resolved by duality transformations and we briefly provide a physical
picture of how this occurs.  The unsafe domains represent true
singularities where all field theoretic description of the physics breaks
down. They violate the holographic principle. We argue that this structure
provides a natural arrow of time for cosmology.  All of the Kasner
solutions, of the compactified SUGRA theory interpolate between the past
and future light cones of the moduli space.  We describe tentative
generalizations of this analysis to moduli spaces with less SUSY.
}

\keywords{M-Theory, String Duality, Superstring Vacua}

\received{???????? ?st, 1998}
\accepted{???????? ?th, 1998}

\preprint{\hepth{9811194}\\RU-98-58, UTTG-13-98\\HEP-UK-0007}
\begin{document}


\section{Introduction}
There have been a large number of papers on the application of string
theory and M-theory to cosmology \cite{cosmoa}-\cite{cosmog}.  In the
present paper 
we will study the cosmology of
toroidally compactified M-theory, and argue that 
some of the singularities encountered in the low energy field theory
approximation can be resolved by U-duality.  We argue that near any of
the singularities we study, new light states appear, which cannot be
described by the low energy field theory.  The matter present
in the original universe decays into these states and the universe
proceeds to expand into a new large geometry which is described by a
different effective field theory.  

In the course of our presentation we will have occasion to investigate the
moduli space of M-theory with up to ten compactified rectilinear toroidal
dimensions and vanishing three form potential.  We believe that the results
of this investigation are extremely interesting.  For tori of dimension, $d
\leq 8$, we find that all noncompact regions of the moduli space can be
mapped into weakly coupled Type II string theory, or 11D SUGRA, at large
volume.  For $d \leq 7$ this result is a subset of the results of Witten
\cite{witten} on toroidal compactification of string theory.  The result
for $ d= 9$ is more interesting.  There we find a region of moduli space
which cannot be mapped into well understood regions.  We argue that a
spacetime cannot be in this regime if it satisfies the Bekenstein bound.

For the ten torus the moduli space can be viewed as a $9+1$ dimensional
Minkowski space.  The interior of the future light cone is the region that
can be mapped into Type II or 11D (we call the region which can be so
mapped the safe domain of the moduli space).  We again argue that the
other regions violate the Bekenstein bound, in the form of the
cosmological holographic principle of \cite{lenwilly}. Interestingly, the
pure gravity Kasner solutions lie precisely on the light cone in moduli
space.  The condition that homogeneous perturbations of the Kasner
solutions by matter lie inside the light cone of moduli space is precisely
that the energy density be positive.

However, every Kasner solution interpolates between the past and future 
light cones.   Thus, M-theory appears to define a natural {\it arrow of 
time} for cosmological solutions in the sense that it is reasonable to 
define the future as that direction in which the universe approaches the
safe domain.  Cosmological solutions appear to  interpolate between a
past where the holographic principle cannot be satisfied and a future
where it can.  

We argue that the $9+1$ dimensional structure, which we derive purely
group theoretically, \footnote{and thus presumably the structure of the
(in)famous hyperbolic algebra $E_{10}$, about which we shall have nothing
to say in this paper, besides what is implicit in our use of its Weyl
group.} is intimately connected to the De~Witt metric on the moduli space.  
In particular, in the low energy interpretation, the signature of the
space is a consequence of the familiar fact that the conformal factor has
a negative kinetic energy in the Einstein action.  Thus, the fact that the
duality group and its moduli space are exact properties of M-theory tells
us that this structure of the low energy effective action has a more
fundamental significance than one might have imagined.  The results of
this section are the core of the paper and the reader of limited patience
should concentrate his attention on them.

In the Section 4 of the paper we speculatively generalize our arguments to
moduli spaces with less SUSY.  We argue that the proper arena for the
study of M-theoretic cosmology is to be found in moduli spaces of compact
ten manifolds with three form potential, preserving some SUSY.  The
duality group is of course the group of discrete isometries of the metric
on moduli space.  We argue that this is always a $p+1$ dimensional Lorentz
manifold, where $p$ is the dimension of the moduli space of appropriate,
static, SUSY preserving solutions of 11D SUGRA in 10 compact dimensions,
restricted to manifolds of unit volume.  We discuss moduli spaces with
varying amounts of SUSY, raise questions about the adequacy of the 11D
SUGRA picture in less supersymmetric cases, and touch on the vexing puzzle
of what it means to speak about a potential on the moduli space in those
cases where SUSY allows one. In the Appendix we discuss how the even
self-dual lattices $\Gamma_8$ and $\Gamma_{9,1}$ appear in our
framework.

\section{Moduli, Vacua, Quantum Cosmology  and Singularities}

\subsection{Some Idiosyncratic Views on General Wisdom}

M-theorists have traditionally occupied themselves with moduli spaces of
Poincar\'e invariant SUSY vacua. It was hoped that the traditional field
theoretic mechanisms for finding stable vacuum states would uniquely pick
out a state which resembled the world we observe.

This point of view is very hard to maintain after the String Duality
Revolution.  It is clear that M-theory has multiparameter families of
exact quantum mechanical SUSY ground states, and that the first
phenomenological question to be answered by M-theory is why we do not live
in one of these SUSY states. It is logical to suppose that the answer to
this question lies in cosmology. That is, the universe is as it is not
because this is the only stable endpoint to evolution conceivable in
M-theory but also because of the details of its early evolution.  To
motivate this proposal, recall that in infinite Poincar\'e invariant space
time of three or more dimensions, moduli define superselection sectors.  
Their dynamics is frozen and solving it consists of minimizing some
effective potential once and for all, or just setting them at some
arbitrary value if there is no potential.  Only in cosmology do the moduli
become real dynamical variables.  Since we are now convinced that
Poincar\'e invariant physics does not destabilize these states, we must
turn to cosmology for an explanation of their absence in the world of
phenomena.

The focus of our cosmological investigations will be the moduli which are
used to parametrize SUSY compactifications of M-theory. We will argue that
there is a natural Born-Oppenheimer approximation to the theory in which
these moduli are the slow variables.  In various semiclassical
approximations the moduli arise as zero modes of fields in compactified
geometries.  One of the central results of String Duality was the
realization that various aspects of the space of moduli could be discussed
(which aspects depend on precisely how much SUSY there is) even in regions
where the notions of geometry, field theory and even weakly coupled string
theory, were invalid.  {\it The notion of the moduli space is more robust
than its origin in the zero modes of fields would lead us to believe.}

The moduli spaces of solutions of the SUGRA equations of motion that
preserve eight or more SUSYs, parametrize exact flat directions of the
effective action of M-theory. Thus they can perform arbitrarily slow
motions. Furthermore, their action is proportional to the volume of the
universe in fundamental units.  Thus, once the universe is even an order
of magnitude larger than the fundamental scale we should be able to treat
the moduli as classical variables.  They provide the natural definition of
a semiclassical time variable which is necessary to the physical
interpretation of a generally covariant theory. In this and the next
section we will concentrate on the case with maximal SUSY.  We relax this
restriction in section 4.  There we also discuss briefly the confusing
situation of four or fewer SUSYs where there can be a potential on the
moduli space.

In this paper we will always use the term moduli in the sense outlined
above. They are the slowest modes in the Born-Oppenheimer approximation to
M-theory which becomes valid in the regime we conventionally describe by
quantum field theory. In this regime they can be thought of as special
modes of fields on a large smooth manifold.  However, we believe that
string duality has provided evidence that the moduli of supersymmetric
compactifications are exact concepts in M-theory, while the field
theoretic (or perturbative string theoretic)  structures from which they
were derived are only approximations.  The Born-Oppenheimer
approximation for the moduli is likely to be valid even in regimes
where field theory breaks down.
The first task in understanding an M-theoretic cosmology is to discuss 
the dynamics of the moduli.  After
that we can begin to ask when and how the conventional picture of quantum
field theory in a classical curved spacetime becomes a good approximation.

\subsection{Quantum Cosmology}

The subject of Quantum Cosmology is quite confusing.  We will try to be
brief in explaining our view of this arcane subject.  There are two issues
involved in quantizing a theory with general covariance or even just time
reparametrization invariance.  The first is the construction of a Hilbert
space of gauge invariant physical states.  The second is describing the
physical interpretation of the states and in particular, the notion of
time evolution in a system whose canonical Hamiltonian has been set equal
to zero as part of the constraints of gauge invariance. The first of these
problems has been solved only in simple systems like first quantized
string theory or Chern-Simons theory, including pure $2+1$ dimensional
Einstein gravity.  However, it is a purely mathematical problem, involving
no interpretational challenges.  We are hopeful that it will be solved in
M-theory, at least in principle, but such a resolution clearly must await
a complete mathematical formulation of the theory.

The answer to the second problem, depends on a semiclassical
approximation.  The principle of time reparametrization invariance forces
us to base our measurements of time on a physical variable.  If all
physical variables are quantum mechanical, one cannot expect the notion of
time to resemble the one we are used to from quantum field theory.  It is
well understood \cite{bfstbruba}-\cite{bfstbrubc} how to derive a
conventional time dependent Schr\"odinger equation for the quantum
variables from the semiclassical approximation to the constraint equations
for a time reparametrization invariant system.  We will review this for
the particular system of maximally SUSY moduli.

In fact, we will restrict our attention to the subspace of moduli space
described by rectilinear tori with vanishing three form.
In the language of 11D SUGRA, we are
discussing metrics of the Kasner form
\eqn{metric}{ds^2 = - dt^2 + L_i^2 (t) (dx^i)^2 }
where the $x^i$ are ten periodic coordinates with period $1$.
When restricted to this class of metrics, the Einstein Lagrangian has
the form
\eqn{lag}{\Lag = V\left[ 
\sum_i {\dot{L}_i^2 \over L_i^2} -
\left(\sum_i {\dot{L}_i \over L_i}\right)^2 
\right],}
where $V$, the volume, is the product of the $L_i$.
In choosing to write the metric in these coordinates, we have lost the
equation of motion obtained by varying the variable $g_{00}$.  This is
easily restored by imposing the constraint of time reparametrization
invariance. The Hamiltonian $E_{00}$ derived from (\ref{lag}) should
vanish on
physical states.  This gives rise to the classical Wheeler-De~Witt equation
\eqn{wd}{
2E_{00} = \left(\sum_i \frac{\dot{L}_i}{L_i}\right)^2 
- \sum_i \left(\frac{\dot{L}_i}{L_i}\right)^2  = 0 ,}
which in turn leads to a naive quantum Wheeler-De~Witt equation:
\eqn{qwd}{{1\over 4V}\left(\sum_i \Pi_i^2 - {2\over 9}(\sum_i \Pi_i)^2 
\right)\Psi = 0.}
That is, we quantize the system by converting the unconstrained phase
space variables (we choose the logarithms of the $L_i$ as canonical
coordinates) 
to operators in a function space.  Then physical states
are functions satisfying the partial differential equation (\ref{qwd}).
There are complicated mathematical questions involved in constructing an
appropriate inner product on the space of solutions, and related
problems of operator ordering.  In more complex systems it is
essential to use the BRST formalism to solve these problems.  
We are unlikely to be able to resolve these questions before discovering
the full nonperturbative formulation of M-theory.   However, for our
present semiclassical considerations these mathematical details are not
crucial. 

We have already emphasized that when the volume of the system is large
compared to the Planck scale, 
the moduli behave classically.  It is then possible to use the time
defined by a particular classical solution (in a particular coordinate
system in which the solution is nonsingular for almost all time).
Mathematically what this means is
that in the large volume limit, the solution to the Wheeler De~Witt
equation takes the form
\eqn{semicsoln}{ \psi_{W\!K\!B} (c) \Psi (q , t[c_0]) }
Here $c$ is shorthand for the variables which are treated by classical
mechanics, $q$ denotes the rest of the variables and $c_0$ is some
function of the classical variables which is a monotonic function of
time.  The wave function $\Psi$ satisfies a time dependent Schr\"odinger 
equation
\eqn{schrod} {i \partial_t \Psi = H(t) \Psi}
and it is easy to find an inner product which makes the space of its
solutions into a Hilbert space and the operators $H(t)$ Hermitian.
In the case where the quantum variables $q$ are quantum fields on the
geometry defined by the classical solution, this approximation is
generally called Quantum Field Theory in Curved Spacetime.  We emphasize
however that the procedure is very general and depends only on the
validity of the WKB approximation for at least one classical variable,
and the fact that the Wheeler De~Witt equation is a second order
hyperbolic PDE, with one timelike coordinate.  These facts are derived
in the low energy approximation of M-theory by SUGRA.  However, we will
present evidence in the next section that they are consequences of the 
U-duality symmetry of the theory and therefore possess a validity beyond
that of the SUGRA approximation.

From the low energy point of view, the hyperbolic nature of the equation
is a consequence of the famous negative sign for the kinetic energy of the
conformal factor in Einstein gravity, and the fact that the kinetic
energies of all the other variables are positive.  It means that the
moduli space has a Lorentzian metric.
 
\subsection{Kasner Singularities and U-Duality}

The classical Wheeler-De~Witt-Einstein equation for Kasner metrics takes
the form:
\eqn{cwde}{({\dot{V} \over V})^2 - \sum_{i=1}^{10}  ({\dot{L_i} \over
L_i})^2 = 0}
This should be supplemented by equations for the ratios of individual
radii $R_i; \prod_{i=1}^{10} R_i =1 $.  The latter take the form of
geodesic motion with friction 
on the manifold of $R_i$ (which we parametrize {\it
e.g.} by the first nine ratios) 

\eqn{nlsigma}{ \partial_t^2 R_i + 
\Gamma^i_{jk} \partial_t R_j  \partial_t R_k + \partial_t ({\rm 
ln} V) \partial_t R_i}
$\Gamma$ is the Christoffel symbol of the metric $G_{ij}$ on the
unimodular moduli space.  We write the equation in this general form
because many of our results remain valid when the rest of the variables 
are restored to the moduli space, and even
generalize to the less supersymmetric moduli
spaces discussed in section 4.  By introducing a new time variable
through
$V(t) \partial_t = - \partial_s$ we convert this equation into
nondissipative geodesic motion on moduli space.  Since the 
``energy'' conjugate to the variable $s$ is conserved, the energy
of
the nonlinear model in cosmic time (the negative term in the Wheeler
De~Witt equation) satisfies
\eqn{en} {\partial_t E = - 2 \partial_t (\mbox{ln\,} V) E}
whose solution is
\eqn{ensol}{ E = {E_0 \over V^2}}
Plugging this into the Wheeler De~Witt equation we find that $V \sim t$ 
(for solutions which
expand as $t \rightarrow \infty$).  Thus, for this class of solutions we
can choose
the volume as the monotonic variable $c_0$ which defines the time in the
quantum theory.

For the Kasner moduli space, we find that the solution of the equations
for 
individual radii
are
\eqn{kassoln}{R_i (t) = \lpl (t/t_0 )^{p_i}}
where
\eqn{kascond}{\sum p_i^2 = \sum p_i = 1}
Note that the equation (\ref{kascond}) implies that at least one of the
$p_i$ is 
negative 
(we have again restricted attention to the case where the volume expands
as 
time goes to 
infinity).    

It is well known that all of these solutions are singular 
at both infinite and zero time.  
Note that if we add a matter or radiation energy density to
the system
then it dominates the system in the infinite volume limit and changes
the 
solutions for the
geometry there.  However, near the singularity at vanishing volume both 
matter and radiation 
become negligible (despite the fact that their densities are 
becoming infinite) 
and the solutions retain their Kasner form.

All of this is true in 11D SUGRA.  In M-theory we know that many regions
of moduli space which are apparently singular in 11D SUGRA can be
reinterpreted as living in large spaces described by weakly coupled Type
II string theory or a dual version of 11D SUGRA. The vacuum Einstein
equations are of course invariant under these U-duality transformations.
So one is lead to believe that many apparent singularities of the Kasner
universes are perfectly innocuous.

Note however that phenomenological matter and radiation densities which
one might add to the equations are not invariant under duality.  The
energy density truly becomes singular as the volume goes to zero.  How
then are we to understand the meaning of the duality symmetry?  The
resolution is as follows.  We know that when radii go to zero, the
effective field theory description of the universe in 11D SUGRA becomes
singular due to the appearance of new low frequency states.  We also know
that the singularity in the energy densities of matter and radiation
implies that scattering cross sections are becoming large.  Thus, it seems
inevitable that phase space considerations will favor the rapid
annihilation of the existing energy densities into the new light degrees
of freedom.  This would be enhanced for Kaluza-Klein like modes, whose
individual energies are becoming large near the singularity.

Thus, near a singularity with a dual interpretation, the contents of the
universe will be rapidly converted into new light modes, which have a
completely different view of what the geometry of space is\footnote{After
this work was substantially complete, we received a paper, \cite{riotto},
which proposes a similar view of certain singularities. See also
\cite{rama}.}. The
most
effective description of the new situation is in terms of the transformed
moduli and the new light degrees of freedom.  The latter can be described
in terms of fields in the reinterpreted geometry.  We want to emphasize
strongly the fact that the moduli do not change in this transformation,
but are merely reinterpreted.  This squares with our notion that they are
exact concepts in M-theory.  By contrast, the fields whose zero modes they
appear to be in a particular semiclassical regime, do not always make
sense.  The momentum modes of one interpretation are brane winding modes
in another and there is no approximate way in which we can consider both
sets of local fields at the same time.  Fortunately, there is also no
regime in which both kinds of modes are at low energy simultaneously, so
in every regime where the time dependence is slow enough to make a low
energy approximation, we can use local field theory.

This mechanism for resolving cosmological singularities leads naturally to
the question of precisely which noncompact regions of moduli space can be
mapped into what we will call the {\it safe domain} in which the theory
can be interpreted as either 11D SUGRA or Type II string theory with radii
large in the appropriate units.  The answer to this question is, we
believe, more interesting than the idea which motivated it.  We now turn
to the study of the moduli space.

\section{The Moduli Space of  M-Theory on Rectangular Tori}

In this section, we will study the structure of the moduli space
of M-theory compactified on various tori $T^k$ with $k\leq 10$.  We
are  especially interested in noncompact regions of this space which
might represent either singularities or large universes. As above, 
the three-form potential $A_{MNP}$ will be
set to zero and the circumferences of the cycles of the torus
will be expressed as the exponentials
\eqn{radiiexp}{ {L_i \over \lpl} = t^{p_i} ,\qquad
i=1,2, \dots, k.}

The remaining coordinates $x^0$ (time) and $x^{k+1}\dots x^{10}$ are
considered to be infinite and we never dualize them.

So the radii are encoded in the logarithms $p_i$. We will study limits of
the moduli space in various directions which correspond to keeping
$p_i$ fixed and sending $t\to\infty$ (the change to $t\to 0$
is equivalent to $p_i\to -p_i$ so we do not need to study
it separately).

We want to emphasize that our discussion of asymptotic domains of moduli
space is complete, even though we restrict ourselves to rectilinear tori
with vanishing three form.  Intuitively this is because the moduli we
leave out are angle variables.  More formally, the full moduli space is
a homogeneous space.  Asymptotic domains of the space correspond to
asymptotic group actions, and these can always be chosen in the Cartan
subalgebra.  The $p_i$ above can be thought of as parametrizing a
particular Cartan direction in $E_{10}$.\footnote{We thank E.Witten for a
discussion of this point.}

\subsection{The \rut}

M-theory has dualities which allows us to identify the vacua with
different $p_i$'s.  A subgroup of this duality group is the $S_k$ which
permutes the $p_i$'s.
Without  loss of generality, we can assume that
$p_1\leq p_2\leq \dots \leq p_9$. We will assume this in most of
the text.
The full group that leaves invariant rectilinear tori with
vanishing three form is the Weyl group of the noncompact $E_k$ group
of SUGRA. We will denote it by $\Rut_k$.  We will give an elementary
derivation of the properties of this group for the convenience of
the reader.  Much of this is review, but our results about the
boundaries of the fundamental domain of the action of $\Rut_k$ with $k =
9,10$ on the moduli space, are new.
$\Rut_k$ is generated
by the permutations, and one other transformation which acts as follows:
\eqn{rutdef}{(p_1,p_2,\dots, p_k)
\mapsto
(p_1-{2s\over 3},
p_2-{2s\over 3},
p_3-{2s\over 3},
p_4+{s\over 3},
\dots,
p_k+{s\over 3}).}
where $s=(p_1+p_2+p_3)$.  
Before explaining why this transformation is a
symmetry of M-theory, let us point out several of its properties
(\ref{rutdef}).

\begin{itemize}
 \item The total sum $S=\sum_{i=1}^k p_i$ changes to $S\mapsto
S+(k-9)s/3$. 
So if $s<0$, the sum increases for $k<9$, decreases for $k>9$
and is left invariant for $k=9$.

 \item If we consider all $p_i$'s to be integers which are
equal modulo 3, this property will hold also after
the \rut. The reason is that, due to the assumptions, $s$ is a multiple
of three and the coefficients $-2/3$ and $+1/3$ differ by an integer.
%
As a result, from any initial integer $p_i$'s we get $p_i$'s
which are multiples of $1/3$ which means that all the matrix elements
of matrices in $\Rut_{k}$ are integer multiples of $1/3$. 

 \item The order of $p_1,p_2,p_3$ is not changed (the difference
$p_1-p_2$ remains constant, for instance). Similarly,
the order of $p_4,p_5,\dots, p_k$ is unchanged. However the
ordering between $p_{1...3}$ and $p_{4...k}$ can change in general.
By convention, we will follow each \rut{} by a permutation which places
the $p_i$'s in ascending order.

\item The bilinear quantity $I= (9 - k) \sum (p_i^2) + (\sum p_i )^2 = (10
- k) \sum(p_i^2) + 
 2 \sum_{i < j} p_i p_j$ is left invariant by $\Rut_k$.
\end{itemize}

The fact that \rut{} is a symmetry of M-theory can be proved as follows.
Let us interpret $L_1$ as the M-theoretical circle of a type IIA string
theory. Then the simplest duality which gives us a theory of the same kind
(IIA) is the double T-duality. Let us perform it on the circles $L_2$
and $L_3$. The claim is that if we combine this double T-duality
with a permutation of $L_2$ and $L_3$ and interpret the new $L_1$ as the
M-theoretical circle again, we get precisely (\ref{rutdef}).

Another illuminating way to view the transformation \rut{} is to
compactify M-theory on a three torus.  The original M2-brane and the
M5-brane wrapped on the three torus are both BPS membranes in eight
dimensions.  One can argue that there is a duality transformation
exchanging them \cite{ofer}. In the limit in which one of the cycles of
the $T^3$ is small, so that a type II string description becomes
appropriate, it is just the double T-duality of the previous paragraph.  
The fact that this transformation plus permutations generates $\Rut_k$ was
proven by the authors of \cite{elitzur} for $k \leq 9$, see also
\cite{pioline}.

\subsection{Extreme Moduli}

There are three types of boundaries of the toroidal moduli space which
are amenable to detailed analysis.  The first is the limit in which
eleven-dimensional supergravity becomes valid. We will
denote this limit as 11D. The other two limits are weakly coupled
type IIA and type IIB theories in 10 dimensions. We will call the domain
of asymptotic moduli space which can be mapped into one of these limits,
the safe domain.

\begin{itemize}

 \item For the limit 11D, all the radii must be greater than $\lpl$.
Note that for $t\to\infty$ it means that all the radii are much greater
than $\lpl$. In terms of the $p_i$'s , this is the inequality $p_i>0$.

 \item For type IIA, the dimensionless coupling constant
$g_s^{IIA}$ must be smaller than 1 (much smaller for $t\to\infty$)
and all the remaining radii must be greater than $\lst$ (much
greater for $t\to\infty$).

 \item For type IIB, the dimensionless coupling constant
$g_s^{IIB}$ must be smaller than 1 (much smaller for $t\to\infty$)
and all the remaining radii must be greater than $\lst$ (much
greater for $t\to\infty$), including the extra radius whose momentum
arises as the number of wrapped M2-branes on the small $T^2$ in the
dual 11D SUGRA picture.

\end{itemize}

If we assume the canonical ordering of the radii, i.e. $p_1\leq p_2\leq
p_3\leq \dots \leq p_k$, we can simplify these requirements as follows:

\begin{itemize}
 \item 11D:    $0<p_1$
 \item IIA:    $p_1<0<p_1+2p_2$
 \item IIB:    $p_1+2p_2<0<p_1+2p_3$
\end{itemize}
To derive this, we have used the familiar relations:
\eqn{fama}{  {L_1\over \lpl}=(g_s^{IIA})^{2/3}=
\left({\lpl\over\lst}\right)^2=
\left({L_1\over \lst}\right)^{2/3}}
for the 11D/IIA duality ($L_1$ is the M-theoretical circle) and similar
relations for the 11D/IIB case ($L_1<L_2$ are the parameters of the
$T^2$ and $L_{IIB}$ is the circumference of the extra circle):
\begin{eqnarray}
{L_1\over L_2}=g_s^{IIB},\quad
1={L_1\lst^2\over\lpl^3}={g_s^{IIB}L_2\lst^2\over\lpl^3}=
{L_{IIB}L_1L_2\over\lpl^3},\\
\frac{1}{g_s^{IIB}}\left(\frac{\lpl}{\lst}
\right)^4=\frac{L_1L_2}{\lpl^2}=\frac{\lpl}{L_{IIB}}=
(g_s^{IIB})^{1/3}\left(\lst\over L_{IIB}\right)^{4/3}\label{famb}
\end{eqnarray}

Note that the regions defined by the inequalities above cannot overlap,
since the regions are defined by $M,M^c\cap A,A^c\cap B$ where
$A^c$ means the complement of a set.
Furthermore, assuming $p_i<p_{i+1}$ it is easy to show that 
$p_1+2p_3<0$ implies $p_1+2p_2<0$ and $p_1+2p_2<0$ implies
$3p_1<0$ or $p_1<0$. 

This means that (neglecting the boundaries where
the inequalities are saturated) the region outside 
$\mbox{11D}\cup\mbox{IIA}\cup\mbox{IIB}$ is defined simply by
$p_1+2p_3<0$.  The latter characterization of the safe domain 
of moduli space will simplify our discussion considerably.

\vspace{3mm}

The invariance of the bilinear form defined above gives an important
constraint on the action of $\Rut_k$ on the moduli space.  For $k=10$ it
is easy to see that, considering the $p_i$ to be the coordinates of a ten
vector, it defines a Lorentzian metric on this ten dimensional space.  
Thus the group $\Rut_{10}$ is a discrete subgroup of $O(1,9)$.  The
direction in this space corresponding to the sum of the $p_i$ is timelike,
while the hyperplane on which this sum vanishes is spacelike. We can
obtain the group $\Rut_9$ from the group $\Rut_{10}$ by taking $p_{10}$ to
infinity and considering only transformations which leave it invariant.  
Obviously then, $\Rut_9$ is a discrete subgroup of the transverse Galilean
group of the infinite momentum frame. For $k \leq 8$ on the other hand,
the bilinear form is positive definite and $\Rut_k$ is contained in
$O(k)$.  Since the latter group is compact, and there is a basis in which
the $\Rut_k$ matrices are all integers divided by $3$, we conclude that in
these cases $\Rut_k$ is a finite group. In a moment we will show that
$\Rut_9$ and {\it a fortiori} $\Rut_{10} $ are infinite. Finally we note
that the \rut{} is a spatial reflection in $O(1,9)$.  Indeed it squares to
$1$ so its determinant is $\pm 1$.  On the other hand, if we take all but
three coordinates very large, then the \rut{} of those coordinates is very
close to the spatial reflection through the plane $p_1 + p_2 + p_3 = 0$,
so it is a reflection of a single spatial coordinate.

\FIGURE[l]{\epsfig{file=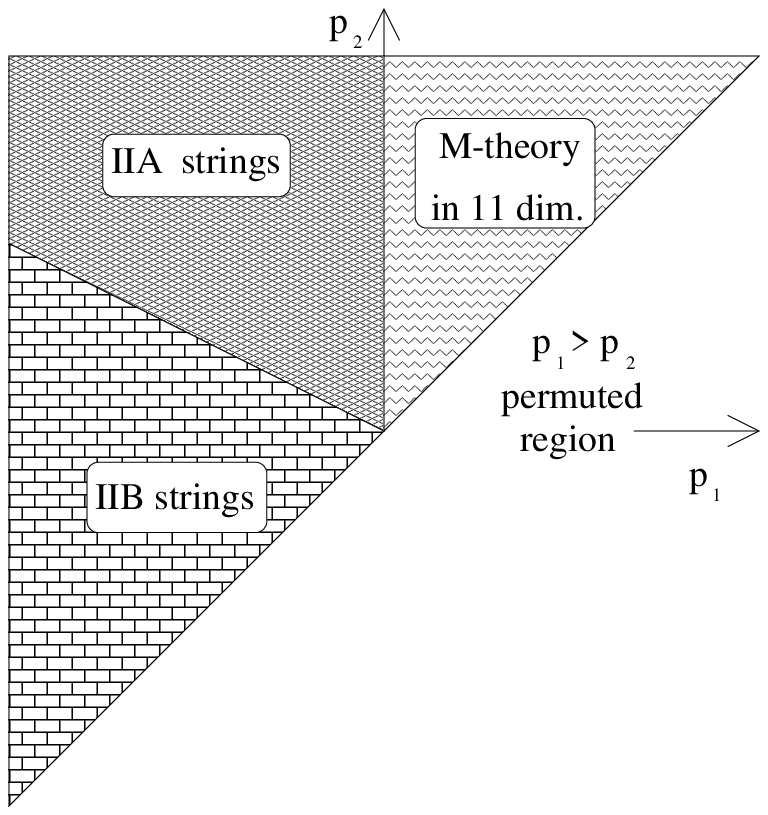}\caption{The structure of the
moduli space for $T^2$.}\label{myfigure}}

\vspace{3mm}

We now prove that $\Rut_9$ is infinite.
Start with the first vector of $p_i$'s given below and iterate
(\ref{rutdef}) on the three smallest radii (a strategy which we will use
all the time) -- and sort $p_i$'s after each step, so that their index
reflects their order on the real line. We get
\eqn{ninefinite}{
\begin{array}{lcr}
(-1,-1,-1,&-1,-1,-1,&-1,-1,-1)\\
(-2,-2,-2,&-2,-2,-2,&+1,+1,+1)\\
(-4,-4,-4,&-1,-1,-1,&+2,+2,+2)\\
(-5,-5,-5,&-2,-2,-2,&+4,+4,+4)\\
{ }&\vdots&{ }\\
(3\times (2-3n),&3\times (-1),&3\times (3n-4))\\
(3\times (1-3n),&3\times (-2),&3\times (3n-2))
\end{array}
}
so the entries grow (linearly) to infinity.

\subsection{Covering the Moduli Space}

We will show that there is a useful strategy which can be used to
transform any point $\{p_i\}$ into the safe domain in the case 
of $T^k$, $k<9$. The strategy
is to perform iteratively \rut{s} on the three smallest radii.

Assuming that $\{p_i\}$ is outside the safe domain, i.e.
$p_1+2p_3<0$ ($p_i$'s are sorted so that $p_i\leq p_{i+1}$),
it is easy to see that $p_1+p_2+p_3<0$ (because $p_2\leq p_3$).
As we said below the equation (\ref{rutdef}), 
the \rut{} on $p_1,p_2,p_3$ always increases the total
sum $\sum p_i$ for $p_1+p_2+p_3<0$. But this sum cannot increase
indefinitely because the group $\Rut_k$ is finite for
$k<9$. Therefore the iteration proccess must terminate at some
point. The only way this can happen is that
the assumption $p_1+2p_3<0$ no longer holds, which means that
we are in the safe domain. This completes the proof for $k<9$.

\vspace{3mm}

For $k=9$ the proof is more difficult. The group $\Rut_9$ is infinite
and furthermore, the sum of all $p_i$'s does not change. In fact 
the conservation of $\sum p_i$ is the reason that only points with
$\sum p_i>0$ can be dualized to the safe domain.
The reason is that if $p_1+2p_3\geq 0$, also $3p_1+6p_3\geq 0$
and consequently
\eqn{ninep}{p_1+p_2+p_3+p_4+p_5+p_6+p_7+p_8+p_9 \geq
p_1 +p_1+p_1 + p_3+p_3+p_3+p_3+p_3+p_3\geq 0.}
This inequality is saturated only if all $p_i$'s are equal
to each other. If their sum vanishes, each $p_i$ must then vanish.
But we cannot obtain a zero vector from a nonzero vector
by \rut{s} because they are nonsingular. If the sum $\sum p_i$ is
negative, it is also clear that we cannot reach the safe domain.

However, if $\sum_{i=1}^9 p_i>0$, then we can map the region of moduli space
with $t \rightarrow \infty $ to the safe domain. 
We will prove it for rational $p_i$'s only. This assumption compensates
for the fact that the order of
$\Rut_9$ is infinite.
Assuming $p_i$'s rational is however sufficient
because we will see that a finite product of \rut{s} brings us
to the safe domain. But a composition of a finite number
of \rut{s} is a continuous map from $\IR^9$ to $\IR^9$ so there must be 
at least a
``ray'' part of a neighborhood which can be also dualized to the
safe domain. Because $\IQ^9$ is dense in $\IR^9$, our argument proves the 
result
for general values of $p_i$.

From now on we assume that the 
$p_i$'s are rational numbers. Everything is scale invariant so we
may multiply them by a common denominator to make integers. In fact, we choose
them to be integer multiples of three since in that case we will have integer
$p_i$'s even after \rut{s}. The numbers $p_i$ are now integers equal
modulo 3 and their sum is positive. We will define a critical quantity
\eqn{cq}{C=\sum_{i<j}^{1...9} (p_i-p_j)^2.}
This is {\it a priori} an integer greater than or equal to zero 
which is invariant
under permutations. What happens to $C$ if we make
a \rut{} on the radii $p_1,p_2,p_3$? The differences
$p_1-p_2$, $p_1-p_3$, $p_2-p_3$ do not change and this holds for
$p_4-p_5$, \dots $p_8-p_9$, too. The only contributions to
(\ref{cq}) which are changed are from $3\cdot 6=18$ ``mixed'' terms like
$(p_1-p_4)^2$. Using (\ref{rutdef}),
\eqn{rutcq}{(p_1-p_4) \mapsto (p_1-\frac{2s}3) -(p_4+\frac{s}3)=
(p_1-p_4)-s}
so its square 
\eqn{rs}{(p_1-p_4)^2\mapsto [(p_1-p_4)-s]^2=(p_1-p_4)^2 - 2s(p_1-p_4)
+s^2}
changes by $- 2s(p_1-p_4) +s^2$. Summing over all 18 terms we get
($s=p_1+p_2+p_3$)
\eqn{delta}{\Delta C= -2s[6(p_1+p_2+p_3)-3(p_4+\dots+p_9)]+18s^2
=6s^2 + 6\left((\sum_{i=1}^9 p_i)-s\right)=6s\sum_{i=1}^9 p_i.}
But this quantity is strictly negative because $\sum p_i$ is positive
and $s<0$ (we define the safe domain with boundaries, $p_1+2p_3\geq 0$).

This means that $C$ defined in (\ref{cq}) decreases after each
\rut{} on the three smallest radii. Since it is a non-negative integer,
it cannot decrease indefinitely. 
Thus the assumption $p_1+2p_3<0$ becomes invalid after a finite number
of steps and we reach the safe domain. 

The mathematical distinction between the two regions of the moduli space
according to the sign of the sum of nine $p_i$'s, has a satisfying
interpretation in terms of the holographic principle.  In the safe
domain, the volume of space grows in the appropriate 
Planck units, while in the region with
negative sum it shrinks to zero.  The holographic principle tells us
that in the former region we are allowed to describe many of the states
of M-theory in terms of effective field theory while in the latter
region we are not.  The two can therefore not be dual to each other.

Now let us turn to the fully compactified case.  As we pointed out, the
bilinear form $I \equiv 2\sum_{i < j} p_i p_j$ defines a Lorentzian
signature metric on the vector space whose components are the $p_i$.  The
\rut{} is a spatial reflection and therefore the group
$\Rut_{10}$ consists of orthochronous Lorentz transformations.  
Now consider a vector in the safe domain.  We can write it as
\eqn{safevec}{(-2, -2 + a_1, 1 + a_2, \ldots,  1+a_9 )S,
\qquad S\in\IR^+}
where the $a_i$ are positive.  It is easy to see that $I$ is positive
on this configuration.  This means that only the inside of the light
cone can be mapped into the safe domain.  Furthermore, since $\sum p_i$
is positive in the safe domain and the transformations are
orthochronous, only the interior of the 
future light cone in moduli space can be mapped
into the safe domain.  

We would now like to show that the entire interior of the forward light
cone can be so mapped.  We use the same strategy of rational coordinates
dense in $\IR^{10}$.  If we start outside the safe domain, the sum of the
first three $p_i$ is negative.  We again pursue the strategy of doing a
\rut{} on the first three coordinates and then reordering
and iterating.  For the case of $\Rut_9$ the sum of the coordinates was
an invariant, but here it decreases under the \rut{} of the three
smallest coordinates, if their sum is negative.  
But $\sum p_i$ is (starting from rational values
and rescaling to get integers congruent modulo three as before) a
positive integer and must remain so after $\Rut_{10}$ operations.  
Thus, after a finite number of iterations, the
assumption that the sum of the three smallest coordinates is negative
must fail, and we are in the safe domain.  In fact, we generically enter
the safe domain before this point.  The complement of the safe domain
always has negative sum of the first three coordinates, but there are
elements in the safe domain where this sum is negative.

It is quite remarkable that the bilinear form $I$ is proportional to 
the Wheeler-De~Witt Hamiltonian for the Kasner solutions:
\eqn{wdI}{\frac{I}{t^2}=\left(\sum_i \frac{dL_i/dt}{L_i}\right)^2
- \sum_i\left(\frac{dL_i/dt}{L_i}\right)^2=\frac{2}{t^2}\sum_{i<j}p_ip_j.}
The solutions themselves thus lie precisely on the future light cone in
moduli space. Each solution has two asymptotic regions ($t \rightarrow
0,\infty$ in (\ref{metric})), one of which is in the past light cone and
the other in the future light cone of moduli space.  The structure of the
modular group thus suggests a natural arrow of time for cosmological
evolution.  The future may be defined as the direction in which the
solution approaches the safe domain of moduli space.  All of the Kasner
solutions then, have a true singularity in their past, which cannot be
removed by duality transformations.

Actually, since the Kasner solutions are on the light cone, which is
the boundary of the safe domain,  we must add
a small homogeneous energy density to the system in order to make this
statement correct.  The condition that we can map into the safe domain
is then the statement that this additional energy density is positive.
Note that in the safe domain, and if the equation of state of this matter
satisfies (but does not saturate) 
the holographic bound of \cite{lenwilly}, this energy density
dominates the evolution of the universe, while near the singularity, it
becomes negligible compared to the Kasner degrees of freedom.  The assumption
of a homogeneous negative energy density is manifestly incompatible with
Einstein's equations in a compact flat universe so we see that the
spacelike domain of moduli space corresponds to a physical situation
which cannot occur in the safe domain.

The backward lightcone of the asymptotic moduli space is, as we have
said, visited by all of the classical solutions of the theory.  
However, it violates the holographic principle of \cite{lenwilly} if we
imagine that the universe has a constant entropy density per comoving
volume.  We emphasize that in this context, entropy means the logarithm
of the dimension of the Hilbert space of those states which can be given
a field theoretic interpretation and thus localized inside the volume.

Thus, there is again a clear physical reason why the unsafe domain of
moduli space  cannot be
mapped into the safe domain.  
Note again that matter obeying the holographic bound of
\cite{lenwilly} in the future, cannot alter the nature of the solutions 
near the true singularities.

\vspace{3mm}

To summarize: the U-duality group $\Rut_{10}$ divides the asymptotic
domains of moduli space into three regions, corresponding to the
spacelike and future and past timelike regimes of a Lorentzian manifold.
Only the future lightcone can be understood in terms of weakly coupled SUGRA or
string theory.  The group theory provides an exact M-theoretic meaning for the
Wheeler-De~Witt Hamiltonian for moduli.  Classical solutions of the low
energy effective equations of motion with positive energy density for
matter distributions lie in the timelike region of moduli space and 
interpolate between the past and future light cones.
We find it remarkable that the purely group theoretical considerations
of this section seem to capture so much of the physics of toroidal
cosmologies.

\section{Moduli Spaces With Less SUSY}

We would like to generalize the above considerations to situations 
which preserve less 
SUSY.  This enterprise immediately raises some questions, the first of which is
what we mean by SUSY.  Cosmologies with compact spatial sections have no
global symmetries in the standard sense
since there is no asymptotic region in which one can define the generators.
We will define a cosmology with a certain amount 
of SUSY by first looking for Euclidean
ten manifolds and three form field configurations which are solutions of the
equations of 11D SUGRA and have a certain number of Killing spinors.
The first approximation to cosmology will 
be to study motion on a moduli space of
such solutions.
The motivation for this is that at least 
in the semiclassical approximation we are guaranteed
to find arbitrarily slow motions of the moduli.  
In fact, in many cases, SUSY
nonrenormalization theorems guarantee that the semiclassical 
approximation becomes valid for
slow motions because the low energy effective Lagrangian of the
moduli is to a large
extent determined by SUSY.  There are however a number of 
pitfalls inherent in our approach.
We know that for some SUSY algebras, the moduli space of 
compactifications to four or six 
dimensions is not a manifold.  New moduli can appear at 
singular points in moduli space
and a new branch of the space, attached to the old one at the 
singular point, must
be added.  There may be cosmologies which traverse from one branch to 
the other in the
course of their evolution.  If that occurs, there will be a point at
which the moduli space approximation breaks down.
Furthermore, there are many examples of SUSY vacua of M-theory which 
have not yet been 
continuously connected on to the 11D limit, even through 
a series of ``conifold'' 
transitions such as those described above \cite{islands}.   
In particular, it has been suggested that there
might be a completely isolated vacuum state of M-theory \cite{evadine}.
Thus it might not be possible to imagine that all cosmological solutions
which preserve a given amount of SUSY are continuously connected to the
11D SUGRA regime.

Despite these potential problems, we think it is worthwhile to 
begin a study of compact, 
SUSY preserving, ten manifolds.  In this paper we will only study 
examples where the
three form field vanishes.  The well known local condition for a 
Killing spinor, 
$D_{\mu} \epsilon = 0$, has as a condition for local integrability 
the vanishing
curvature condition
\eqn{killspin}{R_{\mu\nu}^{ab} \gamma_{ab} \epsilon = 0}
Thus, locally the curvature must lie in a subalgebra of the Lie 
algebra of $Spin (10)$ which
annihilates a spinor.  The global condition is that the holonomy 
around any 
closed path must lie in a subgroup which preserves a spinor.  
Since we are dealing with
11D SUGRA, we always have both the ${\bf 16}$  and ${\bf\overline{16}}$
representations
of 
$Spin (10)$ so SUSYs 
come in pairs.

For maximal SUSY the curvature must vanish identically and the space 
must be a torus.
The next possibility is to preserve half the spinors and this is
achieved 
by manifolds 
of the form $K3 \times T^7$ or orbifolds of them by freely acting
discrete 
symmetries. 

We now jump to the case of 4 SUSYs.   To find examples, it is convenient to
consider the decompositions $Spin (10) \supseteq
 Spin (k) \times Spin (10-k) $.

The ${\bf 16}$ is then a tensor product of two lower dimensional spinors.
For
$k=2$, the holonomy must be contained in $SU(4) \subseteq Spin (8)$ in
order to preserve a spinor, and it then preserves two (four once the
complex conjugate representation is taken into account). The corresponding
manifolds are products of Calabi-Yau fourfolds with two tori, perhaps
identified by the action of a freely acting discrete group.  This moduli
space is closely related to that of F-theory compactifications to four
dimensions with minimal four dimensional SUSY. The three spatial
dimensions are then compactified on a torus. For $k=3$ the holonomy must
be in $G_2 \subseteq Spin (7)$.  The manifolds are, up to discrete
identifications, products of Joyce manifolds and three tori.  For $k=4$
the holonomy is in $SU(2) \times SU(3)$.  The manifolds are free orbifolds
of products of Calabi-Yau threefolds and K3 manifolds.  This moduli space
is that of the heterotic string compactified on a three torus and
Calabi-Yau three fold. The case $k=5$ does not lead to any more examples
with precisely 4 SUSYs.

It is possible that M-theory contains U-duality transformations which map
us between these classes.  For example, there are at least some examples
of F-theory compactifications to four dimensional Minkowski space which
are dual to heterotic compactifications on threefolds.  After further
compactification on three tori we expect to find a map between the $k=2$
and $k=4$ moduli spaces.

To begin the study of the cosmology of these moduli spaces we restrict the
Einstein Lagrangian to metrics of the form 
\eqn{modmet}{ds^2 = - dt^2 +
g_{AB} (t) dx^A dx^B} 
where the euclidean signature metric $g_{AB}$ lies
in one of the moduli spaces.  Since all of these are spaces of solutions
of the Einstein equations they are closed under constant rescaling of the
metric.  Since they are spaces of restricted holonomy, this is the only
Weyl transformation which relates two metrics in a moduli space. Therefore
the equations (\ref{cwde}) and (\ref{nlsigma}) remain valid, where
$G_{ij}$ is now the De~Witt metric on the restricted moduli space of unit
volume metrics.

It is clear that the metric on the full moduli space still has Lorentzian
signature in the SUGRA approximation.  In some of these cases of lower
SUSY, we expect the metric to be corrected in the quantum theory.  
However, we do not expect these corrections to alter the signature of the
metric.  To see this note that each of the cases we have described has a
two torus factor.  If we decompactify the two torus, we expect a low
energy field theoretic description as three dimensional gravity coupled to
scalar fields and we can perform a Weyl transformation so that the
coefficient of the Einstein action is constant.  The scalar fields must
have positive kinetic energy and the Einstein term must have its
conventional sign if the theory is to be unitary.  Thus, the
decompactified moduli space has a positive metric.  In further
compactifying on the two torus, the only new moduli are those contained in
gravity, and the metric on the full moduli space has Lorentzian signature.

Note that as in the case of maximal SUSY, the region of the moduli space
with large ten volume and all other moduli held fixed, is in the future
light cone of any finite point in the moduli space.  Thus we suspect that
much of the general structure that we uncovered in the toroidal moduli
space, will survive in these less supersymmetric settings.

The most serious obstacle to this generalization appears in the case
of 4 (or fewer) supercharges.  In that case, general arguments do not
forbid the appearance of a potential in the Lagrangian for the moduli.
Furthermore, at generic points in the moduli space one would expect
the energy density associated with that potential to be of order the
fundamental scales in the theory.  In such a situation, it is difficult
to justify the Born-Oppenheimer separation between moduli and high
energy degrees of freedom.  Typical motions of the moduli on their
potential have frequencies of the same order as those of 
the ultraviolet degrees of freedom.

We do not really have a good answer to this question.  Perhaps the
approximation only makes sense in regions where the potential is small.
We know that this is true in extreme regions of moduli space in which
SUSYs are approximately restored.
However, it is notoriously difficult to stabilize the system
asymptotically far into such a region.  This difficulty is particularly
vexing in the context of currently popular ideas
\cite{horwita}-\cite{horwitc} in
which the fundamental scale of M-theory is taken to be orders of
magnitude smaller than the Planck scale.

\section{Discussion}

We have demonstrated that the modular group of toroidally compactified
M-theory prescribes a Lorentzian structure in the moduli space which
precisely mirrors that found in the low energy effective Einstein
action.
We argued that a similar structure will be found for moduli
spaces of lower SUSY, although the precise details could not be worked
out because the moduli spaces and metrics on them generally receive
quantum corrections.  As a consequence the mathematical structure of the
modular group is unknown.  Nonetheless we were able to argue that it
will be a group of isometries of a Lorentzian manifold.
Thus, we argue that the generic mathematical structure discussed in   
minisuperspace\footnote{A term which we have avoided up to this point
because it is confusing in a supersymmetric theory.}
approximations to quantum cosmology based on the low
energy field equations actually has an exact meaning in M-theory.
We note however that the detailed structure of the equations will be  
different in M-theory, since the correct minisuperspace is a moduli
space of static, SUSY preserving static solutions.

The Lorentzian structure prescribes a natural arrow of time, with a
general cosmological solution interpolating between a singular past   
where the holographic principle is violated and a future described by
11D SUGRA or weakly coupled string theory where low energy effective
field theory is a good approximation to the gross features of the
universe.   Note that it is {\it not} the naive arrow of time of any given
low
energy gravity theory, which points from small volume to large volume.
Many of the safe regions of moduli space are singular from the point of
view of any given low energy effective theory.  We briefly described how
those singularities are avoided in the presence of matter.

We believe that the connections we have uncovered are important and   
suggest that there are crucial things to be learned from cosmological
M-theory even if we are only interested in microphysics.  We 
realize that we have only made a small beginning in
understanding the import of these observations.  

Finally, we want to emphasize that our identification of moduli spaces
of SUSY preserving static solutions of SUGRA (which perhaps deserve
a more exact, M-theoretical characterization) as the appropriate
arena for early universe cosmology, provides a new starting point for 
investigating old cosmological puzzles.  We hope to report some progress
in the investigation of these modular cosmologies in the near future.

\acknowledgments

We are grateful to Greg Moore and Ed Witten
for valuable discussions. The work of Tom
Banks and Lubo\hacek s{} Motl was supported in part by the DOE under grant
number DE-FG02-96ER40559. The work of Willy Fischler was supported in part
by the Robert Welch Foundation and the NSF under grant number PHY-9219345.

\appendix

\section{The appearance of $\Gamma_8$ and $\Gamma_{9,1}$}

In this appendix we will explain the appearance of the lattices $\Gamma_8$
and $\Gamma_{9,1}$ in our framework. For the
group $\Rut_k$
we have found a bilinear invariant 
\eqn{scpr}{\vec u\cdot \vec v=(9-k)\sum_{i=1}^k (u_iv_i)
+(\sum_{i=1}^k u_i)(\sum_{i=1}^k v_i).}
Now let us take a vector
\eqn{vdef}{\vec v\equiv
\vec v_{123}=(-\frac23,-\frac23,-\frac23,+\frac13,+\frac13,\dots)}
and calculate its scalar product with a vector $\vec p$ according to
(\ref{scpr}). The result is
\eqn{scprp}{\vec v\cdot \vec p=(9-k)\sum_{i=1}^k (v_ip_i)
+\frac{k-9}3 (\sum_{i=1}^k p_i)=(k-9)(p_1+p_2+p_3).}
Thus for $k=9$ the product vanishes and for $k=10$ and $k=8$
the product equals $\pm (p_1+p_2+p_3)$. If the entries $(-2/3)$ are
at the positions $i,j,k$ instead of $1,2,3$, we get $\pm(p_i+p_j+p_k)$.
Substituting $\vec v_{ijk}$ also for $\vec p$, we obtain
\eqn{normv}{\vec v_{ijk}\cdot \vec v_{ijk} = 2(9-k).}
So this squared norm equals $\pm 2$ for $k=8,10$. More generally, we can
calculate the scalar products of any two $v_{ijk}$'s and the result is
\eqn{anysc}{\frac{\vec v_{ijk}\cdot \vec v_{lmn}}{9-k}=\left\lbrace
\begin{array}{ll}
+\frac23+\frac23+\frac23=+2&\mbox{ if $\{i,j,k\}$ and $\{l,m,n\}$ have
3 elements in common}\\
+\frac23+\frac23-\frac13=+1&\mbox{ if $\{i,j,k\}$ and $\{l,m,n\}$ have 
2 elements in common}\\
+\frac23-\frac13-\frac13=0&\mbox{ if $\{i,j,k\}$ and $\{l,m,n\}$ have 
1 element in common}\\
-\frac13-\frac13-\frac13=-1&\mbox{ if $\{i,j,k\}$ and $\{l,m,n\}$ have 
no elements in common}
\end{array}
\right.
}
so the corresponding angles are $0^\circ$, $60^\circ$, $90^\circ$
and $120^\circ$.
Now a reflection with respect to the hyperplane perpendicular to
$\vec v\equiv \vec v_{ijk}$
is given by
\eqn{reflp}{\vec p\mapsto \vec p' = \vec p-2\vec v\,\,\frac{\vec
p\cdot\vec
v}{\vec v\cdot\vec v}}
and we see that for $k\ne 9$ it precisely reproduces the 2/5
transformation (\ref{rutdef}).

Let us define the lattice $\Lambda_k$ to be the lattice of all integer
combinations of the vectors $\vec v_{ijk}$. We will concentrate on the
cases $k=8$ and $k=10$ since $\Lambda_k$ will be shown to be even
self-dual lattices.
Thus they are isometric to the unique even self-dual lattices in these
dimensions.

It is easy to see that for $k=8,10$, the lattice $\Lambda_k$ contains
exactly those
vectors whose entries are multiples of $1/3$ 
equal modulo 1.
The reason is that all possible multiples of $1/3$ modulo 1
i.e. $-1/3,0,+1/3$ are realized by $-\vec v_{ijk},0,\vec v_{ijk}$
and we can also change any coordinate by one (or any integer) because
for instance
\eqn{byone}{\begin{array}{rcll}
\vec v_{123}+\vec v_{456}+\vec v_{789}&=&(0,0,0,0,0,0,
0,0,0,1)&\mbox{for }k=10\\
\vec v_{812}+\vec v_{345}+\vec v_{678}&=&(0,0,0,0,0,0,
0,-1)&\mbox{for }k=8
\end{array}}
Since the scalar products of any two $\vec v_{ijk}$ are integers
and the squared norms of $\vec v$'s are even, $\Lambda_8$ and
$\Lambda_{10}$ are even lattices. Finally we prove that they are
self-dual.

The dual lattice is defined as the lattice of all vectors whose scalar
products with any elements of the original lattice 
(or with any of its generators $\vec v_{ijk}$) are integers. Because of
(\ref{scprp}) it means that the sum of any three coordinates should be
an integer. But it is easy to see that this condition is identical to the
condition that the entries are multiples of $1/3$ equal modulo 1.
The reason is that the difference
$(p_1+p_2+p_3)-(p_1+p_2+p_4)=p_3-p_4$ must be also an integer -- so
all the coordinates are equal modulo one. But if they are equal modulo 1,
they must be equal to a multiple of $1/3$ because the sum of three such
numbers must be integer.

We think that this construction of $\Gamma_8$ and $\Gamma_{9,1}$ is 
more natural in the
context of U-dualities than the construction using the root lattice of
$E_8$ with the $SO(16)$ sublattice generated by $\pm e_i\pm e_j$.

\vspace{3mm}

Let us finally mention that the orders of $\Rut_3,$
$\Rut_4,$\dots $\Rut_8$ are equal to $2\cdot 3!$,
$5\cdot 4!$, $16\cdot 5!$, $72\cdot 6!$, $576\cdot 7!$ and
$17280\cdot 8!$. For instance the group $\Rut_4$ is isomorphic
to $S_5$. We can see it explicitly. Defining
\eqn{fourfromfive}{p_i=R_i-\frac 13(R_1+R_2+R_3+R_4-R_5),\qquad
i=1,2,3,4}
which leaves $p_i$ invariant under the transformation
$R_i\mapsto R_i+\lambda$, the permutation of $R_4$ and $R_5$ is easily
seen to generate the 2/5 transformation on $p_1,p_2,p_3$. Note that
$p_1+p_2+p_3=-R_4+R_5$.




\begin{thebibliography}{19}        
\bibitem{cosmoa} G.\,Veneziano, 
{``Scale Factor Duality For Classical And Quantum Strings,''}
\plb{265}{1991}{287-294}.
\bibitem{cosmovf} A.A.\,Tseytlin, C.\,Vafa,
{``Elements of String Cosmology,''} \npb{372}{1992}{443-466},
\hepth{9109048}. 
\bibitem{cosmob} M.\,Gasperini, G.\,Veneziano,
{``Pre -- Big Bang In String Cosmology,''}
{\it Astropart. Phys. }{\bf 1} (1993) 317-339.
\bibitem{cosmobtwo} A.\,Buonanno, T.\,Damour, G.\,Veneziano,
{``Prebig Bang Bubbles From The Gravitational Instability Of Generic
String Vacua,''} \hepth{9806230} and references therein;
see also {\tt http://www.to.infn.it/{}$\widetilde{\,\,\,}$gasperin/}.
\bibitem{cosmoc} P.\,Binetruy, M.K.\,Gaillard,
{``Candidates For The Inflaton Field in Superstring Models,''}
\prd{34}{1986}{3069-3083}.
\bibitem{cosmomoore} G.\,Moore, J.H.\,Horne,
{``Chaotic Coupling Constants,''}
\npb{432}{1994}{109-126}, \hepth{9403058}.
\bibitem{cosmod} T.\,Banks, M.\,Berkooz, S.H.\,Shenker,
G.\,Moore, P.J.\,Steinhardt,
{``Modular Cosmology,''}, \prd{52}{1995}{3548-3562}, \hepth{9503114}.
\bibitem{cosmoe} T.\,Banks, M.\,Berkooz, P.J.\,Steinhardt,
{``The Cosmological Moduli Problem, Supersymmetry Breaking, And Stability
In Postinflationary Cosmology,''} \prd{52}{1995}{705-716},
\hepth{9501053}.
\bibitem{cosmofold} A.\,Lukas, B.A.\,Ovrut, {``U Duality Covariant
M~Theory Cosmology,''} \plb{437}{1998}{291-302}, \hepth{9709030},
and references therein.
\bibitem{cosmof} S.-J.\,Rey, D.\,Bak, 
{``Holographic Principle and String Cosmology,''}
\hepth{9811008}.
\bibitem{cosmog} O.\,Bertolami, R.\,Schiappa,
{``Modular Quantum Cosmology,''} \grqc{9810013}.
%
\bibitem{witten} E.\,Witten, {``String Theory Dynamics In Various
                 Dimensions,''} \npb{443}{1995}{85-126}, \hepth{9503124}.
\bibitem{lenwilly} W.\,Fischler, L.\,Susskind,
                 {``Holography and Cosmology,''} \hepth{9806039}.
\bibitem{bfstbruba} T.\,Banks, W.\,Fischler, L.\,Susskind,
  {``Quantum Cosmology In $(2+1)$-Dimensions And $(3+1)$-Dimensions,''}
  \npb{262}{1985}{159}.
\bibitem{bfstbrubb} T.\,Banks, {``TCP, Quantum Gravity, The Cosmological
Constant And All That\dots,''} \npb{249}{1985}{332}.
\bibitem{bfstbrubc} V.A.\,Rubakov, 
{``Quantum Mechanics In The Tunneling Universe,''}
\plb{148}{1984}{280-286}.
\bibitem{riotto} M.\,Maggiore, A.\,Riotto, {``D-Branes and Cosmology,''}
\hepth{9811089}.
\bibitem{rama} S.K.\,Rama, {``Can String Theory Avoid Cosmological
Singularities?''} \plb{408}{1997}{91-97}, \hepth{9701154}.
\bibitem{ofer} O.\,Aharony, 
{``String Theory Dualities from M-Theory,''}
\npb{476}{1996}{470-483}, \hepth{9604103}.
\bibitem{elitzur} S.\,Elitzur, A.\,Giveon, D.\,Kutasov, E.\,Rabinovici,
{``Algebraic Aspects of Matrix Theory on $T^d$,''}
\npb{509}{1998}{122-144}, \hepth{9707217}.
\bibitem{pioline} N.A.\,Obers, B.\,Pioline,
{``U-Duality and M-Theory,''} 
invited review for {\it Phys. Rept,}\newline
\hepth{9809039}.
\bibitem{islands} A.\,Dabholkar, J.A.\,Harvey,
{``String Islands,''} \hepth{9809122}.
\bibitem{evadine} M.\,Dine, E.\,Silverstein,
{``New M-theory Backgrounds with Frozen Moduli,''}\newline
\hepth{9712166}.
\bibitem{horwita} P.\,Ho\hacek rava, E.\,Witten,
{``Heterotic and Type I String Dynamics from Eleven Dimensions,''}
\npb{460}{1996}{506-524}, \hepth{9510209}.
\bibitem{horwitb} E.\,Witten, {``Strong Coupling Expansion Of Calabi-Yau
Compactification,''} \npb{471}{1996}{135-158}, \hepth{9602070}.
\bibitem{horwitc} T.\,Banks, M.\,Dine, 
{``Couplings And Scales In Strongly Coupled Heterotic String Theory,}
\npb{479}{1996}{173-196}, \hepth{9605136}.
\bibitem{horwitd} N.\,Arkani-Hamed, S.\,Dimopoulos, G.\,Dvali, 
{``The Hierarchy Problem and New Dimensions at a Millimeter,''}
\plb{429}{1998}{263-272}, \hepph{9803315}.
%
\end{thebibliography}
\end{document}